\documentclass[aps,twocolumn,prl,showpacs]{revtex4} 

\usepackage{amsmath}  
\usepackage{psfrag}  
\usepackage{graphicx} 

\usepackage{psfig} 
\usepackage{color}  

\newcommand{\Vect}[1]{\boldsymbol{\rm #1}} 
\newcommand{\Tens}[1]{\boldsymbol{\rm #1}}

\DeclareMathOperator{\Tr}{Tr}  
\newcommand{\Jump}[1]{\left[#1\right]^{\text{out}}_{\text{in}}} 
   
\newcommand{\Vnabla}{\Vect{\nabla}}

\newcommand{\Vr}{\Vect{r}}  
   
\newcommand{\Vv}{\Vect{v}}  
\newcommand{\Vro}{\Vect{R}}   
  
\newcommand{\Vn}{\Vect{n}}

\begin{document}   

\title{Wrinkling of microcapsules in shear flow}    
\author{Reimar Finken}% \email{reimar.finken@theo2.physik.uni-stuttgart.de}  
\author{Udo Seifert} 
\affiliation{
II. Institut f{\"u}r Theoretische Physik,
Universit{\"a}t Stuttgart, 
70550 Stuttgart, 
Germany}

%\date{\today}      

\begin{abstract} 
  Elastic capsules can exhibit short wavelength wrinkling in external shear
  flow. We analyse this
  instability  of the capsule shape and use the length scale
  separation between the capsule radius and the wrinkling wavelength to derive
  analytical results both for the threshold value of the shear rate and for
  the critical wave-length of the wrinkling. These results can be used to
  deduce elastic parameters from experiments.
\end{abstract}  
% insert suggested PACS numbers in braces on next line  \pacs{} 

% \keywords{}   
\maketitle  
% body of paper here 
% 
% General outline: 
% - Introduction 
% - locally Cartesian coordinate system -> strain and curvature 
% - wrinkling as 3d perturbation 
% - elastic response 
% - hydrodynamic response 
% - small time-limit 
% - application  

The pronounced shape transformations of vesicles and microcapsules induced by
hydrodynamic flow paradigmatically illustrate a main theme in microfluidics.
The dynamical balance between fluid-induced viscous and membrane-determined
elastic stresses depends crucially and distinctively on the specific soft
object immersed into the flow. In shear flow, {\sl fluid} bilayer vesicles
assume a stationary tank-treading shape if there is no viscosity contrast
between the interior and exterior fluid \cite{kraus1996}. If the interior
fluid or the membrane gets more viscous a transition to a tumbling state can
occur \cite{biben2003a,beaucourt2004b,rioual2004a,noguchi2004a,noguchi2005a,vitkova2005a}. While comprehensive
experiments are still lacking, tank-treading has been
observed both for vesicles in infinite shear flow \cite{haas1997} and, in
particular, for vesicles interacting with a rigid wall
\cite{lorz2000,abka02} where a dynamical lift occurs
\cite{seifert1999b,cantat1999a,sukumaran2001,beaucourt2004a}. For a
comprehensive review of the dynamical behaviour of soft capsules in shear flow,
we refer the reader to the first two chapters of \cite{pozrikidis2003book}. 

\begin{figure}    
\centering   
\includegraphics[width=0.5\linewidth]{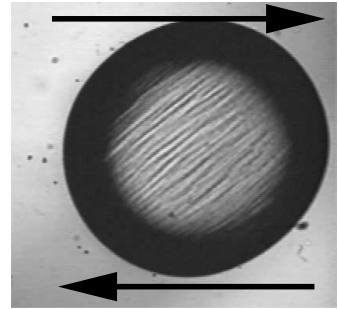}   
\caption{Wrinkling of a polysiloxane capsule ($\simeq 343 \mu {\rm m}$) in shear
  flow as indicated by the arrows (adapted from \cite{walter2001}).}   
\label{fig:wrinkling}  
\end{figure}

In contrast to fluid vesicles, genuine microcapsules exhibit a finite shear
rigidity since their membrane is physically or chemically cross-linked. While
shear elasticity prevents very large shape transformations such as a
prolate-oblate transition in fluid vesicles \cite{noguchi2004a}, it also leads
to qualitatively different instabilities in shear flow like the wrinkling
instability first observed experimentally \cite{walter2001}, see Fig.
\ref{fig:wrinkling}. In fact, wrinkling is a ubiquitous phenomenon which can
occur whenever a compressive force acts on a thin sheet either clamped at
boundaries or attached to an elastic substrate as happens for the wrinkling of
skin \cite{cerda2002,cerda2003}.

The quantitative study of wrinkling is challenging since the usually large
deformations of thin sheets lead to nonlinear partial differential equations
for the force balance. Still, rather general statements about wrinkling
phenomena become possible, since the nonlinearity is mostly due to geometric
effects and thus universal. Indeed, it has been pointed out by Cerda {\em et
  al.} in a series of intriguing papers \cite{cerda2002,cerda2003,cerda2004b}
that geometry puts such a strong constraint on the underlying physics, that
wrinkles which form in very different contexts all obey the same scaling laws.
They consequently identified as the necessary ingredients for all wrinkling
phenomena a thin sheet with a bending stiffness, an effective elastic
foundation, and an imposed compressive strain \cite{cerda2003}. These
ingredients give rise to a rather general effective Hamiltonian description,
where long wavelength wrinkles are punished energetically due to the
increasing longitudinal stretching strain, while short wavelength wrinkles
increase the bending energy. This leads to the selection of the optimal
wrinkling wavelength.

% *************************************************

Since Cerda {\em et al.} considered universal scaling behaviour of all
wrinkling phenomena, conversely no conclusions concerning the physical origin
underlying the effective constants can be drawn from the scaling behaviour
alone. Considering the elastic capsule system in particular, identifying the
effective elastic stiffness of the rather complex coupled fluid-membrane
system is not trivial. Moreover, the Hamiltonian description is \emph{a
  priori} no longer valid when the compressive forces acting on the membrane
are nonconservative in nature, e.g. when arising from a \emph{hydrodynamic}
flow. Here, dynamical effects have to be taken into account. In this letter we
therefore complement the approach discussed above by taking advantage of the
short wavelength of the observed wrinkles. We can thus identify a new
mechanism by which short wavelength wrinkles are selected in a {\em curved
  geometry}. This approach can also be generalised to the full dynamic
problem, although we have so far only taken into account the short-time
dynamics. Moreover, the exact asymptotic limit allows us to
make quantitative (i.e. non scaling argument) predictions on the onset and
direction of wrinkles.

Assuming linear elasticity (and presupposing the capsule was initially
stress-free) the work required to deform the membrane is given by the
expression
\begin{equation}    
\label{eq:1}   
\mathcal{H}[\Vro] = \frac{1}{2}\int dA\left(\lambda  \Tr(\Tens{\epsilon})^{2} +
  2 \mu \Tens{\epsilon} \mathbf{\colon} \Tens{\epsilon}  +   \kappa \left(2H-\frac{2}{R}\right)^{2}\right).  
\end{equation} 
Here $\lambda$ is the two-dimensional isotropic compressibility, $\mu$ the
Lam{\'e} coefficient corresponding to the shear stress, and $\kappa$ the
bending modulus of the elastic material, for which we assume for simplicity a
spontaneous curvature given by the initial capsule radius. The resulting forces can be calculated
as functional derivatives of the energy with respect to the deformation
$\Vect{f}^{\text{el}} = - \delta \mathcal{H}[\Vect{R}] /\delta \Vect{R}$. 
% \begin{align}    
% \label{eq:4}   
% \Vect{f}^{\text{el}} \equiv - \frac{\delta
%   \mathcal{H}[\Vect{R}]}{\delta \Vect{R}} =  \Vnabla \cdot \left[\lambda  \Tr(\Tens{\epsilon}) \Tens{1} + 2 \mu  \Tens{\epsilon} + \kappa \left(2H-\frac{2}{R}\right) \Tens{K} \right] + \Delta \left(2H-\frac{2}{R}\right)  \Vect{n}   
% \end{align}  
We now immerse this capsule of radius $R$ into a fluid of
viscosity $\eta$  subject to the unperturbed shear flow
\begin{equation}   
\label{eq:5}       
\Vv^{\infty} = \dot{\gamma}[(y/2)\Vect{e}_{x}+(x/2)\Vect{e}_{y}],
\end{equation} 
where $\dot{\gamma}$ denotes the shear rate. (Here $x$ and $y$ refer to global
Cartesian coordinates fixed in a laboratory frame.) Considering more general
linear shear flow is straightforward, but would distract from the wrinkling
problem. For the system at hand the resulting Reynolds number of the flow is
negligible, so that the velocity field $\Vect{v}$ and pressure $p$ are
governed by the Stokes equation $\eta \Delta \Vect{v} = \Vnabla p$ together
with the incompressible continuity equation $\Vnabla \cdot \Vect{v} = 0$. With
the definition of the hydrodynamic stress $\Tens{\sigma}= - p \Tens{1} + \eta
\left[ \Vnabla \Vect{v} + (\Vnabla \Vect{v})^{T}\right]$ the Stokes equation
can also be written as $\Vnabla \cdot \Tens{\sigma} = 0$. The hydrodynamic
force acting on the membrane is given by $\Vect{f}^{\text{fl}} \equiv
\Jump{\Tens{\sigma}\cdot \Vn}$. The Stokes equations are accompanied by the
asymptotic boundary condition $\Vect{v}(\Vect{r}) \rightarrow
\Vect{v}^{\infty}$ for $|\Vect{r}|\rightarrow \infty$, the force balance
$\Vect{f}^{\text{el}} + \Vect{f}^{\text{fl}} = 0$ at the membrane, and the
kinematic condition that the membrane is convected by the fluid,
$\left.\Vect{v} \right|_{\Vect{R}(\theta,\phi,t)} = \partial_{t}
\Vect{R}(\theta,\phi,t)$.

These equations fully determine the evolution of the capsule shape.
Unfortunately they are highly nonlinear and non-local, so that a solution is
only possible numerically or in certain asymptotic limits. Let us now suppose
that we have found a stationary solution, and look for the stability of small
perturbations $\delta \Vr$ of the membrane position. From the experimental
evidence \cite{walter2001} it is expected that beyond a critical shear rate
$\dot{\gamma}_{c}$ the stationary deformation becomes unstable with respect to
the formation of short wavelength wrinkles. For a first qualitative
discussion, we consider a given configuration with known initial strain tensor
$\Tens{\epsilon}_{0}$ and impose oscillatory wrinkles perpendicular to the
membrane with the large wavevector $k \Vect{e}_{x}$, leading to $\delta \Vr =
\delta r \cos(kx) \Vect{e}_{z}$. In this one-dimensional picture, the strain
is simply the relative change in length of a line element with respect to a
reference element. An initially straight line (a cut through a \emph{flat}
membrane) always increases in length when folded. The relative length change
is of the order $\delta l_{1} / l \sim (\delta r)^{2} k^{2}$. This effect is
of second second order in the amplitude and grows quadratically with the
wavenumber. For a \emph{curved} membrane, there is an additional first order
length change $\delta l_{2} / l \sim \delta r \cos(k x)/ R$, arising from the
fact that the local curvature radius of the line is periodically displaced
outward and inward. Both contributions are added to the initial strain. For
the \emph{curved} membrane, the relevant component of the strain tensor thus
changes to
\begin{equation}
  \label{eq:13}
  \epsilon \sim \epsilon_{0} + (\delta r)^{2} k^{2} + \delta r \cos(kx)/R.  
\end{equation}
Even though the curvature induced part is oscillatory and vanishes in the
mean, it nevertheless gives rise to a positive contribution in the elastic
energy, which is quadratic in the strain. For large wavenumbers, the local
curvature, which is of the order $k^{2} \delta R$, gives rise to the most
important contribution to the total energy. After averaging over the fast
oscillations, one obtains for the elastic energy (ignoring pre-factors of the
order one and the shear elasticity for a moment)
\begin{equation}
  \label{eq:24}
 \mathcal{H}^{\text{el}} \sim \mathcal{H}^{\text{el}}_{0} + \int dA 
       \frac{\lambda}{R^{2}} (\delta r)^{2} + \lambda \epsilon_{0} (\delta r)^{2} k^{2}  + \kappa
       (\delta r)^{2}  k^{4}.
\end{equation}
This expression has formally the same form as the effective Hamiltonian in
\cite{cerda2003}, so that we can identify the bending stiffness $\kappa$, the
stress $\lambda \epsilon_{0}$, and the effective elastic foundation stiffness
$\lambda / R^{2}$. Before we discuss the latter quantity in more detail, let
us briefly recapitulate how these terms lead to short wavelength wrinkling for
sufficiently large compressive stress: For an initially compressed membrane
(i.e. $\epsilon_{0} < 0$), the compressive stress is partially canceled by the
second order contribution $(\delta r)^{2} k^{2}$. This lowers the modulus of
the resulting stress and therefore decreases energy. This mechanism is also
the driving force for Euler buckling of planar membranes and rods. For a
curved membrane, there is an additional contribution that always serves to
increase the elastic energy. Only with a sufficiently large compressive
pre-strain ($\epsilon_{0} < 0$) one is able to overcome this energy cost.
Since the energy gain is quadratic in the wavenumber, short wavelength
wrinkles will overcome the cost associated with the curvature effect sooner
for increasing $|-\epsilon_{0}|$. However, very large wavenumbers are
penalised by the bending energy. A simple discussion of the functional form
\eqref{eq:24} thus reveals that gaining elastic energy by wrinkling is
possible provided that the initial compressive stress exceeds a threshold
$\tau_{0} = \lambda |\epsilon_{0}| > \tau_{c} \sim \sqrt{\lambda \kappa}/ R$.
Even though this scaling prediction is already implicit in \cite{cerda2003}
the $1/R$ term shows the crucial influence of a curved geometry here.  

This critical stress must be provided by the imposed hydrodynamic flow. From a
simple dimensional analysis one can deduce that this stress is of the order
$\tau \sim \eta \dot{\gamma} R$. For the critical stress we find a critical
shear rate $\dot{\gamma}_{c} \sim \sqrt{\kappa \lambda}/(\eta R^{2})$ and a
critical wrinkling wavenumber $k_{c} \sim (\lambda/(\kappa R^{2}))^{1/4} \sim
1/(hR)^{1/2}$, where the second relation arises when we treat the membrane as
an isotropic shell of thickness $h$ and bulk modulus $E$, i.e. $\lambda \sim E
h$ and $\kappa_{c} \sim E h^{3}$.

This discussion reveals the physical origin of the effective stiffness
$\lambda / R^{2}$: Any normal deformation on a \emph{curved} membrane leads to
an additional strain contribution that increases the elastic energy to second
order in the amplitude. This picture becomes more subtle when we allow the
membrane to relax tangentially: In plane displacements will reduce the second
order strain to minimise its energy contribution. The two dimensional freedom
in general does not allow to reduce the full stress tensor, which consists of
three independent components, to zero. This becomes possible, however, for
particular values of the material constants or the membrane geometry. If the
shear modulus of the membrane vanishes, tangential deformations can completely
eliminate the isotropic strain. In such a membrane the effective stiffness
consequently vanishes. This explains why wrinkling is not observed for fluid
membranes. When the curvature perpendicular to the wrinkle wavevector
vanishes, it becomes again possible to eliminate the total additional stress.
Thus the wrinkling phenomenon on a capsule necessarily needs both a curved
geometry and finite values for the compressibility and shear modulus.

The quantitative treatment is considerably more complicated. We not only have
to express the change in the elastic energy due to wrinkling not necessarily
perpendicular to the membrane on a quantitative level, but we also have to
take the change in the hydrodynamic forces due to the deformed capsule into
account, since change in the forcing linear in the amplitude acts on the same
level as the quadratic change of the elastic energy. However, we can neglect
all these complications if we assume that the number of wrinkles on the
membrane, $k_{c} R$, is large. Formally, we are looking for the singular
perturbation solution of the dynamic equations in the limit $\varepsilon
\equiv (\kappa/(\lambda R^{2}))^{1/4} \sim 1/(k_{c}R)
\rightarrow 0$.
When we compare the elastic forces with the change in the hydrodynamic forces
in detail, we we find that the latter can be ignored as being of the order $\varepsilon$.  

\begin{figure}
  \centering
      \includegraphics[width=\linewidth]{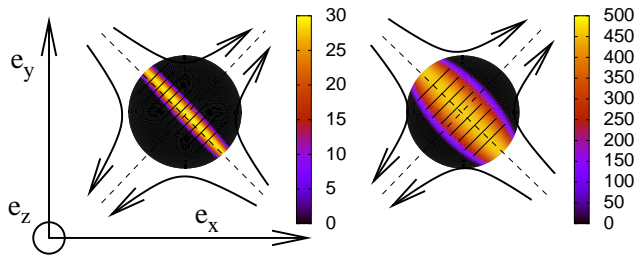}
    \caption{Wrinkling instability region for $\dot{\gamma}=11.5 {\rm s^{-1}}$ (left)
      and $\dot{\gamma}=17.0 {\rm s^{-1}}$ (right), respectively.
        The
      elastic moduli of the membrane are $\lambda = \mu = 0.1 \mathrm{N
        m^{-1}}$, $\kappa = 1\cdot 10^{-17} \mathrm{Nm}$. In the
      black colored region the membrane is still stable with respect to
      wrinkle formations. Brighter colors indicate a larger growth rate of the
      wrinkles. Initially wrinkling occurs only in a small region around the
      circumference corresponding to the plane of maximal compressive stress.
      At higher shear rates this region grows, as does the growth rate of each
      wrinkling mode. The direction of the wrinkles along the lines of maximal
      tension are indicated schematically.}
   \label{fig:region}
\end{figure}

We now turn to the \emph{quantitative} predictions. The wrinkling pattern is
assumed to be of the more general form $\delta \Vr = \delta \Vr_{0} \exp(i
\Theta)$, where the local wavevector is now given as the gradient of the phase
$\Vect{k} \equiv \Vnabla \Theta$. The amplitude $\delta \Vr_{0}$ and the
wavevector $\Vect{k}$ are allowed to vary slowly on the membrane. The
collection of the most important contributions to the change in elastic energy
is formally obtained via WKB theory \cite{hinch1991}. We need to consider
non-normal deformations of the membrane, which allows the membrane to reduce
the stress even further. We will publish the details of the calculation
elsewhere and only note that the energy change due to the formation of
wrinkles on a sphere is
\begin{equation}    
\label{eq:14}   
 \delta^{2} \mathcal{H} =  \frac{1}{4}  \int dA \left\{ 4 \mu \frac{\lambda +
     \mu}{\lambda + 2 \mu} \frac{1}{R^{2}} + \Vect{k} \cdot \Tens{\tau} \cdot \Vect{k} + \kappa |k|^{4}\right\}\delta r^{2}.
\end{equation}
Here $\Tens{\tau}$ denotes the tangential part of the elastic stress tensor,
which is effectively a symmetric $2 \times 2$ matrix determined by the
external forces. When this matrix has a sufficiently large negative
eigenvalue, energy can be gained from the formation of wrinkles. The force
driving the folding is the negative derivative of \eqref{eq:14} with respect
to $\delta r$ and in this order linear in the amplitude. The elastic forces in
turn induce a flow in the surrounding fluid that convect the membrane points.
In general, this effect would be non-local, leading to a coupling between the
wrinkling dynamics at different points of the capsule. However, at zero order
in the asymptotic short wavelength limit the hydrodynamic effects are local
and diagonal in the coordinate system adapted to the membrane. We are thus led
to the evolution equation
\begin{equation}
  \label{eq:12}
 \partial_{t}  (\delta r) = - \frac{1}{8 \eta |k|} 
 \left\{ \frac{4 \mu }{R^{2}}\frac{\lambda +
     \mu}{\lambda + 2 \mu} + \Vect{k} \cdot \Tens{\tau} \cdot \Vect{k} +
   \kappa |k|^{4}\right\}\delta r \equiv \frac{\delta r}{\tau},
\end{equation}
which implies a time scale $\tau$ for the short time dynamics of each
wrinkling mode. We take the
time constant as a measure of the strength of the instability and assume
 that locally the fastest growing mode is dynamically selected. We can
thus find the resulting folding pattern by maximising the r.h.s. of equation \eqref{eq:12}. Note that
the only input needed for this theory is the initial stress distribution
$\Tens{\tau}$, which is caused by the hydrodynamic flow. For general shapes,
these data can come e.g. from numerical simulations such as
\cite{noguchi2004a}. For the present purpose, we will use analytic results for
the expansion of the stress to first order in the shear rate $\dot{\gamma}$
\cite{barthes-biesel1980}. The stress distribution on a stationary sphere in
linear shear flow \eqref{eq:5} is independent of the material constitutive law
and reads in spherical coordinates, see Fig. \ref{fig:region},
\begin{equation}
  \label{eq:19}
  \Tens{\tau} =
  \begin{pmatrix}
    \tau_{\theta\theta} & \tau_{\theta\phi}\\
    \tau_{\phi\theta} & \tau_{\phi\phi}
  \end{pmatrix} = \dot{\gamma} \frac{5}{2} \eta R
  \begin{pmatrix}
     \sin 2\phi &  \cos 2\phi \cos \theta\\
     \cos 2\phi \cos \theta &  - \sin 2\phi  \cos^{2} \theta 
  \end{pmatrix}.
\end{equation}
The wavevector field of the wrinkles is given by the direction of the negative
eigenvalues of \eqref{eq:19}. Even though these can be calculated analytically, they are
given by rather involved expressions. However, one can see that the most
negative stress eigenvalue $- 5 \dot{\gamma} \eta R / 2$ appears along the
circumference $\phi = 3 \pi / 4, 7 \pi /4$, which is at $\pi/4$ angle with the
shear direction. The corresponding eigenvector
points perpendicular to the this circumference along the $\theta$ coordinate lines. Along this circumference the local
compression is maximal, from which we can calculate the critical shear rate $\dot{\gamma}_{c}$
 and the corresponding critical wavenumber $k_{c}$ analytically as
\begin{equation}
  \label{eq:21}
  \dot{\gamma}_{c}  = \frac{8}{5 \eta R^{2}} \sqrt{\mu
      \kappa \frac{\lambda+\mu}{\lambda + 2 \mu}} = \frac{4 \kappa}{5 \eta R} k_{c}^{2} 
\end{equation}
These two expressions are our main quantitative results. The scaling is of
course the same as predicted in \cite{cerda2003}. However, we also predict
quantitative prefactors that become exact in the large wavenumber limit. Above
the critical shear rate $\dot{\gamma}_{c}$ wrinkles start to form in a finite
region starting uniformly along the circumference $\phi = 3 \pi /4, 7 \pi /4$.
With increasing shear rates this wrinkling region grows symmetrically towards
the stagnation points of the flow located at $\theta = \pi /2$, $\phi = \pi
/4, 5 \pi /4$. The wrinkling direction is determined by the direction of the
eigenvectors of the stress tensor. Along the initial wrinkling circumference,
the folds form perpendicular to this line, i.e. in the direction of maximal
tension, in accordance with experimental findings \cite{walter2001}.

Inserting experimental value for the planar elastic moduli \cite{walter2001}
$\lambda \simeq \mu \simeq 0.1\, {\rm Nm^{-1}}$ and counting the number of
wrinkles across the membrane (c.f. Figure \ref{fig:wrinkling}) we arrive at an
estimate for the bending rigidity $\kappa \simeq 1\cdot 10^{-17}\,{ \rm Nm}$,
in accordance with a similar estimate made in \cite{cerda2003}. Of course, due
to the fourth power taken, the estimation for $\kappa$ (c.f.
equation~\eqref{eq:21}) depends sensitively on the measured wavenumber. For
these values, together with the capsule radius $R\simeq 343 \, \mu {\rm m}$
and the viscosity of water, $\eta = 10^{-3}\,{\rm Pa \cdot s}$, the critical
shear rate is predicted to be $\dot{\gamma}_{c} \sim 11 {\rm s}^{-1}$. This
value is in good agreement with the experimentally reported onset at
$\dot{\gamma}_{c} \simeq 4 {\rm s}^{-1}$ \cite{walter2001}. For shear rates
larger than $\dot{\gamma}_{c}$, the minimization of \eqref{eq:12} is best
performed numerically. The growth of the wrinkling region for the numerical
shear rates $\dot{\gamma} = 11.5 {\rm s}^{-1}$ and $\dot{\gamma} = 17.0 {\rm
  s}^{-1}$ is shown in Figure
\ref{fig:region}.

In conclusion, we have presented an analytical theory for the wrinkling
instability of a microcapsule in shear flow. This was accomplished by a
systematic singular perturbation expansion in the bending rigidity. The only
input needed, apart from the material constants, is the initial stress
distribution. For quasispherical capsules all quantities can be calculated
explicitly, leading to analytical expressions of the shear rate and the
wrinkling wavelength at the onset. This provides a method to deduce the
bending rigidity from shear experiments. Given the relatively large error in
the experimental values of the compressibility the agreement of the predicted
critical shear rate (deduced from the observed wavelength) with experimental
values is surprisingly good.
 
Finally we remark that an initial stress on the membrane (e.g. due to osmotic
pressure) affects the critical shear rate. While the effect of membrane
prestress on the deformation of capsules has been discussed \cite{lac2005},
the effect on the formation of wrinkles has not yet been considered. Since
wrinkling can only occur for membranes with shear stress resistance, fluid
membrane capsules do not form wrinkles. Thus a study of viscoelastic
membranes, which exhibit fluid or elastic behaviour depending on the
timescale, seems worthwhile.

Financial support of the DFG within the priority program ``Nano- and
Microfluidic'' is acknowledged.

\end{document}